\documentclass[11pt]{cernrep}
\usepackage{graphicx}
\usepackage{here}
\def\re#1{(\ref{#1})}
\def\beq{\begin{equation}}
\def\eeq{\end{equation}}
\def\beeq{\begin{eqnarray}}
\def\beeqn{\begin{eqnarray*}}
\def\eeeq{\end{eqnarray}}
\def\eeeqn{\end{eqnarray*}}

\newcommand{\OO}{{\cal O}}

\def\de{\delta}

\def\l{\lambda}                 
\def\m{\mu}
\def\n{\nu}

\def\s{\sigma}                  
\def\th{\theta}

\def\z{\zeta} 

\newcommand{\lp}{\left(}
\newcommand{\rp}{\right)}
\renewcommand{\lq}{\left[}
\renewcommand{\rq}{\right]}

\newcommand{\no}{\nonumber}

\def\tr{\,\mbox{Tr}\,}
\def\frac#1#2{ {{#1} \over {#2} }}

\def\half{\mbox{\small $\frac{1}{2}$}}
\def\p{\partial}

\def\ie{\hbox{\it i.e.}{ }}

\begin{document}
\title{PLANARITY IN NONCOMMUTATIVE GAUGE THEORIES \footnote{\quad
Talk presented by
A. Bassetto at the Light-Cone workshop LC03 ``Hadrons and beyond'', Durham\, (U.K.), 
August 2003.}}
\author{Antonio Bassetto\,\, \S\quad and Federica Vian \,\,\P }
\institute{\S\quad Dipartimento di Fisica ``G. Galilei'', via Marzolo 8, 35131 Padova (Italy)
and Istituto Nazionale di Fisica Nucleare,
Sezione di Padova, Italy\\
\P\quad Associata alla
Presidenza dell'Istituto 
Nazionale di
Fisica Nucleare, Italy, and NORDITA, Blegdamsvej 17, 2100 Copenhagen \O \, (Denmark)}
\maketitle
\begin{abstract}
Planarity was introduced by 't Hooft in his topological classification of
diagrams in the large-$N$ limit of $U(N)$ gauge theories. Planarity also occurs
in noncommutative field theories where amplitudes possess invariance only under 
cyclic permutations, a feature inherited from the parent string theory. 
In noncommutative
gauge theories both kinds of planarity merge in a context which turns out to be 
particularly intriguing in the two-dimensional case where gauge invariant correlators 
can be explicitly computed.
\end{abstract}

\section{INTRODUCTION}

Planarity is an important concept in usual Quantum Field Theories (QFT), especially
in gauge theories (GT), after the seminal work by 't Hooft \cite{thoo},  
who showed that in $U(N)$ GT a topological classification of diagrams is possible
in the large-$N$ limit at a fixed value of $g^2 N$, $g$ being the coupling constant.
Such a classification leads to a power expansion of amplitudes in the variable
$1/N$, the planar diagrams providing the leading contribution.
He also introduced a quite convenient double-line notation for diagrams, each
(oriented) line being associated to a fundamental representation, in which
topology turns out to be particularly transparent.

\smallskip

On a different side, 
in recent years interest has grown for QFT defined on noncommuting space-time
variables, mainly triggered by the work of Seiberg and Witten \cite{sw}, who 
related them to suitable particular limits of a string theory
in the presence of a constant background. In such theories amplitudes are
invariant only with respect to cyclic permutations of external
momenta, a feature inherited from the parent string. In a perturbative
formulation the only difference with the usual Feynman expressions is the
occurrence in the vertices
of a phase factor (Moyal phase), depending on the momenta and
on the noncommutativity parameter $\theta$. Such a phase
affects non-planar diagrams, thereby leading
again to a topological classification \cite{ms}. In turn the presence of a
Moyal phase in general provides a damping factor in the large-$\theta$
limit.

\smallskip

Noncommutative $U(N)$ GT exhibits both kinds of planarities; it is our
purpose to further elaborate and elucidate this point. In so doing we
follow essentially ref.\cite{bv}. 
The merging of space-time and ``internal'' symmetries does not come
out unexpected in such a context; as a matter of fact, when noncommutative
GT are represented in a separable Hilbert space, the gauge
group embodying the mentioned symmetries turns out to be
the set of all unitary operators of the kind ${\cal U}= {\bf I}+{\cal K}$,
${\cal K}$ being suitable compact operators \cite{ha} \footnote{This does not
contradict the well known Coleman-Mandula theorem, as the required hypotheses
are not fulfilled in this instance. One of us (A.B.) wishes to thank T. Heinzl
for calling his attention on this point}. To clarify this merging is one
of the most fascinating and intriguing challenges in our opinion.

In Sect.2 essential definitions and notations of noncommutative
GT are presented; observable
quantities, in particular the open Wilson lines, which are the concrete tool
of our subsequent investigation, are introduced. 
Qualitative consequences of planarity for 
open Wilson line correlators in four (space-time) dimensions are also
exhibited. In Sect.3 correlators are explicitly computed in two dimensions,
both by a perturbative and by a non-perturbative approach, 
in a suitable 
region of the external variables (the ``planar'' phase of the theory),
and concluding remarks are drawn.

\section{Observables in noncommutative gauge theories}

\subsection{Notations and definitions}

Noncommutativity of $D$-dimensional Minkowski 
space-time is encoded in a real antisymmetric matrix 
$\theta^{\mu\nu}$:
\begin{equation}
\label{alge}
[x^\mu,x^\nu]=i\theta^{\mu\nu},\quad\quad\quad\quad \mu,\nu=0,..,D-1,
\end{equation}
and a $\star$-product of two fields $\phi_1(x)$ and $\phi_2(y)$  can be defined
by means of Weyl symbols
\begin{equation}
\label{star}
\phi_1(x)\star\phi_2(y)=\int \frac{d^Dp\, d^Dq}{(2\pi)^{2D}}\exp \lq -\frac{i}2 \, 
p_{\mu}\theta^{\mu\nu}q_{\nu}\rq \exp[i(px+qy)] \tilde \phi_1(p) \tilde \phi_2(q).
\end{equation}
Then noncommutative theories are most easily formulated by replacing the 
usual multiplication of fields in the Lagrangian with the
$\star$-product. The resulting action makes them obviously non-local.

\smallskip

The classical action of the $U(N)$ Yang-Mills theory in a noncommuting
space-time is
\begin{equation}
\label{act}
S=-\frac12 \int d^Dx\, Tr F_{\mu\nu} \star  F^{\mu\nu}\,,
\end{equation}
where the field strength $F_{\m\n}$ is given by
\begin{equation}
F_{\mu\nu}=\partial_{\mu} A_\n -\p_\n A_\m -ig (A_\m\star A_\n - A_\n\star A_\m)
\end{equation}
and $A_\m=A_\m^a T^a$ is a $N\times N$ matrix, with $T^a$ normalized
as follows: $Tr T^a T^b=\half \de^{ab}$, $a,b$ denoting $U(N)$ indices.

The action eq.~\re{act} is invariant under infinitesimal $U(N)$
noncommutative gauge transformations 
\begin{equation}
\label{gauge}
\de_\l A_\m= \p_\m \l -i g (A_\m\star\l -\l\star A_\m) \,.
\end{equation}
As noticed in \cite{gross}, under this transformation 
the operator $Tr F^2(x)$ is not left invariant 
\begin{equation}\label{opera}
\tr F^2(x) \longrightarrow \tr U(x) \star  F^2(x) \star U^\dagger (x)\,,
\end{equation}
with $U(x)=\exp_* (ig \l (x))$. To recover a gauge invariant operator,
one has to integrate over the entire space-time, since  
$\star$-products  inside integrals can be cyclically permuted.
 
As a consequence, gauge invariance in this case (star-gauge invariance) 
entails an integration over space-time variables and the
possibility of having {\it local} probes is lost.

A Wilson line of length $l$ can be  defined by
means of the Moyal product as \cite{gross}
\beq
\label{wline}
\Omega_\star[x,C]=P_{\star} \exp \lp
ig\int_0^l A_\m 
(x+\z(\s))\, d\z^\m(\s)\rp \,,
\eeq
where $C$ is the curve 
parameterized by $\z(\s)$, with $0 \leq \s \leq 1$, $\z(0)=0$,
$\z(1)=l$, and $P_\star$
denotes noncommutative path ordering along $\z(\s)$ from right to left
with respect to increasing $\s$ of $\star$-products of functions.
The Wilson line is not invariant under a gauge transformation
\beq\label{linetrans}
\Omega_\star [x, C] \longrightarrow  U(x) \star \Omega_\star [x,C] \star 
U^\dagger (x+l)\,.
\eeq

The following operator
\beq \label{linedef0}
W(p, C)= \int d^Dx \, \tr \Omega_\star[x, C]\star e^{ipx}\,,
\eeq
turns out to be invariant provided $C$ satisfies the condition
\beq\label{endpoints}
l^\n=p_\m \th^{\m\n}
\eeq
(the Wilson line extends in the direction transverse to the momentum).
The particular case $p_{\mu}=0$ corresponds to a closed loop.

For simplicity in the following only straight lines will be considered.
Then one can easily realize that any local operator $\OO (x)$ in ordinary
gauge theories admits a noncommutative generalization
\beq \label{ope}
{\tilde \OO (p)}= 
\tr \int d^Dx \, \OO (x) \star \Omega_\star [x, C] 
\star e^{ipx}\,,
\eeq
each of the ${\tilde \OO (p)}$'s being a genuinely different operator
at different momentum.

Remarkably, owing to eq.~\re{endpoints}, at large values of $|p|$, 
gauge invariance requires that
the length of the Wilson line  becomes large. This feature can be interpreted
as a manifestation of the UV-IR mixing phenomenon.

\subsection{Two open-line correlator}

An interesting  quantity to study is the two-point function
$\langle W(p)W^\dagger (p)\rangle$, where $W(p)$ has been
defined via eqs.~\re{wline}, \re{linedef0}. It represents the
correlation function of two straight parallel Wilson lines of equal
length, each carrying a transverse momentum $p$.
In four dimensions such a correlator
was investigated in \cite{gross}, according to the perturbative expansion
\beeq \label{pertline}
&&W(p)\,
=\,
\sum_{j=0}^\infty \,(ig)^j \int d^4x\int_{\z_j>\z_{j-1}>\ldots>\z_1}
[d\z]\, \tr
A(x+\z_1) \star  \ldots \star A(x+\z_j)\star e^{ipx}
\no\\
&&W^\dagger(p)\,
=\,
\sum_{j=0}^\infty \,(-ig)^j \int d^4x
\int_{\z'_j>\z'_{j-1}>\ldots>\z'_1} [d\z']\, \tr
A(x+\z'_j)  \star \ldots \star A(x+\z'_1)\star
e^{-ipx}\,.\no\\
&&
\eeeq
By resumming {\it ladder} diagrams, the
correlator was found to grow exponentially at large momenta
\begin{equation}
\label{grrr}
\langle W(p)W^\dagger (p)\rangle \propto \exp \sqrt{\frac{g^2 N|p\theta||p|}
{4\pi}}.
\eeq
This was correctly interpreted in \cite{gross} as a coherence effect, increasing with the
length of the (parallel) lines. Ladder diagrams are leading at large $N$
and their Moyal phases cancel, so their $\theta$ dependence only occurs
in the length of the line. They are {\it planar} according to both ``colour''
and ``geometry'' criteria.

Imagine we now perform a cyclic permutation on one of the lines in a ladder
diagram; the colour
factor is obviously unchanged (each line entails an independent trace
over colour matrices). As a consequence the diagram remains leading as far
as colour is concerned. However the $\theta$-dependence is not insensitive
to such a permutation: Moyal phases double instead of cancelling, producing,
on an intuitive basis, a damping for large values of $\theta$. This is indeed
confirmed by explicit low-order calculations. Diagrams which would be planar
according to colour, do not according to geometry. Planarity in four dimensions
just means ``ladder''.

\section{TWO-LINE CORRELATOR IN TWO DIMENSIONS}

\subsection{A perturbative approach}

In two dimensions the situation is quite different: noncommutativity involves
the time variable, but the Lorentz symmetry is not violated owing to the
tensorial character of $\theta^{\mu\nu}= \theta \epsilon^{\mu\nu}$.
Invariance under area-preserving diffeomorphisms is preserved as well.
If we choose the light-cone gauge, perturbative calculations are greatly 
simplified, thanks to the decoupling of Faddeev-Popov ghosts
and to the vanishing of the vector vertices. 
It turns out that in all  diagrams 
contributing to the line correlators, which are planar according to
the 't Hooft's large-$N$ limit, 
$\th$-dependent phases resulting from non commutativity play no role.
They are {\it planar} also according to ``geometry''. This feature,
which is characteristic of the theory in two dimensions, might be related
to its invariance under area-preserving diffeomorphisms.

\smallskip

On the other hand, in two dimensions a remarkable symmetry, the Morita
equivalence, allows the mapping of open Wilson lines on a noncommutative
torus onto closed Wilson loops winding on a dual commutative torus \cite{sch}.
In turn, in a commutative setting, Wilson loop correlations can be obtained by
geometrical techniques \cite{mig}; this opens the possibility of confronting perturbative
calculations with non-perturbative solutions, provided a common kinematical region
of validity is found for both approaches.

We quantize the theory in the light-cone gauge $A_-=0$ at {\em equal
times}, the free propagator having the following causal expression
(WML prescription)
\begin{equation}
\label{WMLprop}
D^{WML}_{++}(x)={1\over {2\pi}}\,\frac{x^{-}}{-x^{+}+i\epsilon x^{-}}\,,
\end{equation}
first proposed by T.T. Wu \cite{wu}. This propagator is nothing
but the restriction in two dimensions of the expression proposed 
by S. Mandelstam and G. Leibbrandt \cite{ml} in four
dimensions  and 
derived by means of a canonical quantization in \cite{bl}.
It allows a smooth transition to an Euclidean 
formulation, where momentum integrals are performed by means of a
``symmetric integration'' \cite{wu}.

We go back to eq.~\re{pertline} and,
with no loss of generality thanks to
the persisting boost invariance,
we choose the path
$C$ stretching along $x^0$, so that $p$ points in the $x^1$ direction.

We then contract the $A$'s in such a way that the resulting diagram
is of leading order in $N$, which yields, according to eq.~\re{WMLprop},
at a fixed perturbative
order $(g^2)^n$ 
\beq \label{1graph}
(-1)^{n-k}\lp\frac{N}{4\pi}\rp^n 
\int [d\s]\, [d\s']\int d^2x \,e^{ipx}\prod_{j=1}^k
\frac{x^0+f_j(\s,\s',\th p)
-x^1}{-x^0-f_j(\s,\s',\th p)-x^1}\,,\qquad n\ge 1,
\eeq
where $k$ is the number of propagators connecting the two lines~\footnote
{The contribution of propagators starting and ending on the same line
factorizes and amounts to $\lp -N/(4\pi) \rp^{n-k}$.}   and
$f_j(\s,\s',\th p)$ is a linear function of its variables
depending  on the topology;
the integration region for the $2n$ geometric
variables is understood and the phase factors containing the
noncommutativity parameter have been absorbed in the function
$f_j(\s,\s',\th p)$. 

We stress that, remarkably, factorization of propagators in coordinate
variables  occurs just in those diagrams which are dominant at
large $N$. This feature in turn makes $\th$-dependence trivial, as it was
explicitly shown in \cite{bv}, since it intervenes just through the length
of the line $l$.

Surprisingly enough, it will turn out that all
integrals will give the same result, no matter what the function
$f_j(\s,\s',\th p)$ is, \ie no matter what topology we choose in the
set of planar diagrams.
This finding is a direct consequence of the integration over the
world volume $d^2x$, required by noncommutative gauge invariance,
and of the orthogonality of the momentum with respect to the direction of
the open lines. In so doing the $\theta$-dependence is washed out
apart from its occurrence in $l^0$.

In order to provide a correct formulation of the theory, 
continuation to Euclidean variables is required: $x^0 \to i x_2$;
we recall that, to keep the basic algebra unchanged, the 
noncommutativity parameter $\theta$ has also
to be simultaneously continued to an imaginary value: $\theta \to i \theta$.

A symmetric integration \cite{wu}
then provides the natural regularization in eq.~\re{1graph}
\beq\label{x1int}
\int d^2x \, e^{-ip\cdot x} \prod_{j=1}^k
\frac{x_1+i(f_j(\s,\s',\th p)+x_2)}{x_1-i(f_j(\s,\s',\th p)+x_2)}
=(-1)^k\frac{4\pi k}{p^2}.
\eeq
Hence the integration over the geometrical variables in eq.~\re{1graph} is
straightforwardly carried out and yields 
\beq \label{ordern}
\frac{4 \pi k}{p^2} \lp\frac{Nl^{2}}{4\pi}\rp^n
\frac1{n_1!\,n_2!}\,,
\eeq
where $l=|p\th|$ is the total length of the line and  $n_1$, $n_2$ are 
the number of legs stretching out of the first and the second line,
respectively ($n_1+n_2=2n$). 

Eq.~\re{ordern} displays a trivial dependence on the
topology of the graph, the only remnant being ${n_1!\,n_2!}$ in the
denominator. Thus, although resumming even only leading contributions
in $N$ may have seemed a formidable task when we started, it has now
become feasible, provided the exact number of different configurations 
with fixed $n_1$, $n_2$ is known.
A careful counting of all such configurations has been performed in \cite{bv}.
By eventually completely resumming the perturbative series with the
appropriate weight factors included, we obtain the expression

\beeq \label{completata}
&&\langle W(p)W^\dagger (p)\rangle=\frac{4\pi \tau^2}{p^2}\lq
I_0(2\tau)+
\int_{\gamma-i\infty}^{\gamma+i\infty}\ \frac{dz}{8\pi i}\frac{z+\sqrt{z^2-4}}
{\sqrt{z^2-4}} e^{z\tau}\right. \\ \no
&& \times \left(\sqrt{1+(z-\sqrt{z^2-4})^2}-1-\frac12 (z-\sqrt{z^2-4})^2
\right)  \left(z+\sqrt{z^2-4}+2\tau \right)\\ \no
&&+
\int_{\nu_1-i\infty}^{\nu_1+i\infty} \int_{\nu_2-i\infty}^{\nu_2+i\infty}
\frac{dz\ dw}{(2\pi i)^2}  e^{(z+w)\tau} \frac{z^3w^3(1+zw)}{4(zw-1)^3}
\left(1+\frac{2}{z^2}-\sqrt{1+\frac{4}{z^2}}\right)\\ \no
&&\left. \times \left(1+\frac{2}{w^2}-\sqrt{1+\frac{4}{w^2}}\right)
\rq
\,,
\eeeq
where $\tau=\sqrt{\frac{g^2 N l^2}{4\pi}}$,
 $\nu_1,\nu_2>1$ and $\gamma> 2 $.

One can realize that, at large $\tau$, the term with the double integration
dominates and $\langle W(p)W^\dagger (p)\rangle$ 
increases like $\exp (2\tau)=\exp(\sqrt{g^2 N l^2/\pi})$, disregarding a
(small) power correction.
As expected, the correlator depends on the 't Hooft coupling $\sqrt{g^2 N}$;
what is remarkable is that  its asymptotics is an exponential linearly 
increasing with the line momentum $|p|$. This is reminiscent of what was found 
in \cite{gross} for its four-dimensional analog.

It is tempting to argue that for a general correlator of an arbitrary  number
of open parallel lines in two dimensions, $\th$-dependence is trivial,
intervening 
only through the length of the lines, just in those diagrams  which
are dominant at large $N$.  
Indeed, in ref.~\cite{pol} this statement was proved at least for the
correlators of  three parallel Wilson lines. 
In the large-$N$ planar
limit, the perturbative series was resummed. Although multiple line
correlators in a generic configuration are not expected to increase
with the length of the lines on the basis of the estimate in
\cite{gross}, it was found instead they keep increasing, when the
lines are parallel, at the same rate as the two-line
correlator, so that the normalized three-line correlator is still
increasing like a (small) power of its argument. Actually, the
interference effect generated by lines with 
the same orientation is overwhelmed by the coherent increase due to
parallelism of lines with opposite orientation.

\subsection{An approach based on the Morita equivalence}

We turn now our attention to a non-perturbative
derivation of the noncommutative Wilson lines correlator, and see how
it compares with eq.~\re{completata} at large $l$. To this regard,
it is worth noticing 
that although eq.~\re{completata} follows from a perturbative analysis,
having resummed all orders, it holds also at
large $g^2 N$.

When both coordinates are compactified to form a torus, 
a remarkable symmetry, called Morita equivalence \cite{sch}, relates different
noncommutative gauge theories living on different noncommutative tori: the
duality group $SO(2,2,{\bf Z})$ has an $SL(2,{\bf Z})$ 
subgroup which acts as follows 
\begin{equation}
\label{morita1}
\left( \begin{array}{c}
m' \\
N'
\end{array} \right) =
\left( \begin{array}{cc}
a & b \\
c & d
\end{array} \right)
\left( \begin{array}{c}
m \\
N
\end{array} \right), \qquad \Theta'=\frac{c+d\Theta}{a+b\Theta},
\end{equation}
\begin{equation}
\label{morita2}
(R')^2=R^2 (a+b\Theta)^2,\qquad  (g')^2=g^2|a+b\Theta|, \qquad \tilde \Phi'=
(a+b\Theta)^2 \tilde \Phi-b(a+b\Theta),
\end{equation}
where $\Theta \equiv \theta/(2\pi R^2)$, $\tilde \Phi \equiv 2\pi R^2\Phi$, $\Phi$
being a background connection and $R$ the radius of the torus, which, for
simplicity, we assume to be square. The first entry $m$ denotes the
magnetic flux, while $N$ characterizes the gauge group $U(N)$. It is
not restrictive to consider the quantities $m$ and $\theta$ to be
positive. The parameters
of the transformation are integers, constrained by the condition
$ad-bc=1$.

The map in the equations above is flexible enough  to allow
for a commutative theory on the second torus by choosing $\Theta'=0$.
As a consequence, the parameter $d$ will be set equal to $-c/\Theta$,
$\Theta$  being a suitable rational quantity. 
In the sequel, for notational convenience, all the primed quantities
will be affected the subscript $c$ ($N'\equiv N_c,m'\equiv m_c,...$). 

We are eventually 
interested in a noncommutative
theory defined on a plane $(R \to \infty)$, with a trivial first Chern 
class ($m=0,\Phi=0$) and a gauge group $U(N)$ with a large $N$, since we want to
establish a comparison with the perturbative approach of the previous subsection.

We start by considering the action
\beq
\label{action}
S=\frac{1}{4g_c^2}\int d^2 x \,  \tr\left[\left(F_{\mu\nu}-\frac{m_c}{2\pi R_c^2 N_c}
\epsilon_{\mu\nu} {\bf I}\right)\left(F^{\mu\nu}-\frac{m_c}{2\pi R_c^2 N_c}
\epsilon^{\mu\nu} {\bf I}\right)\right],
\eeq
where the explicit expression for the background connection 
$\Phi_c=-\frac{m_c}{2\pi R_c^2 N_c} {\bf I}$ has been introduced.

The formula for the partition function on a torus reads \cite{mig}
\beq
\label{migdal}
{\cal Z}=\sum_{{\cal R}} \exp\left[-\frac{{\cal A}}{2}C_2({\cal R})\right],
\eeq
$C_2$ being the second Casimir operator in the representation ${\cal R}$ and
${\cal A}=4\pi^2 (g_c R_c)^2$.

After performing a harmonic analysis, retaining only the contribution of 
the $m_c$-th sector and cancelling the $U(1)$ contribution
against the background connection, we obtain the final expression
\beeq
\label{final}
{\cal Z}&=&\sqrt{\frac{2\pi}{{\cal A}N_c}}\frac{1}{N_c !}\sum_{n_i\neq n_j}
\exp\left[-\frac{{\cal A}}{2}\left(\sum_{i=1}^{N_c} n_i^2-\frac{1}{N_c}
\left(\sum_{i=1}^{N_c}n_i\right)^2\right)\right]\\ \nonumber
&\times&\int_0^{2\pi}\frac{d\alpha}{\sqrt {\pi}} 
\exp\left[-\left(\alpha -\frac{2\pi}{N_c}
\sum_{i=1}^{N_c}n_i\right)^2 -2\pi im_c\left(
\frac{N_c-1}{2}- \frac{1}{N_c}
\sum_{i=1}^{N_c}n_i\right)\right].
\eeeq

Now we turn our attention to the 
correlation function of two straight parallel Wilson lines of equal length,
lying on the noncommutative torus without winding around it, each carrying
a transverse momentum $p$. The noncommutative torus will eventually
be decompacted by sending its radius $R\to \infty$.

On the noncommutative torus, with the line $C$ stretching along $x_2$, 
we have the expression
\beq
\label{linet}
W(k, C)=\frac{1}{4\pi^2 R^2}\int_0^{2\pi R} d^2 x \, \tr \Omega_\star 
[x,C]\star \exp(i k x_1/R
),
\eeq
where $k$ is the integer associated to the transverse momentum $p=\frac{k}{R}$.
The no-winding condition entails the constraint $\theta k<2\pi R^2$, namely
$l=p \theta<2\pi R$, $l$ being the total length of the straight line. 
$W(k)$ is normalized according to $W(0)=1$.

Now we exploit again the Morita equivalence in order to map the open Wilson line
on the noncommutative torus on a closed Polyakov loop of the ordinary Yang-Mills theory
winding $k$ times around the commutative torus in the $x_2$ direction
\beeq
\label{map}
W(k)&=&W^{(k)},\\ \nonumber
W^{(k)}&=&\frac{1}{4\pi^2 R_c^2}\int_0^{2\pi R_c} d^2x \frac{1}{N_c}
\tr\left[\Omega^{(k)}
(x_1)\right].
\eeeq
The trace is to be taken in the fundamental representation of $U(N_c)$ and
$\Omega^{(k)}(x_1)$ is the holonomy of the closed path. This
holonomy is to be computed in the flux sector $m_c$.

The correlation function of two straight parallel open Wilson lines
reads 
\beeq
\label{corre}
&&{\cal W}_2(k)\equiv <W(k)W(-k)>\\  \nonumber
&=&\ \frac{1}{2\pi R_c} 
\int_0^{2\pi R_c} dx <\frac{1}{N_c} \tr\left[
\Omega^{(k)}(x)\right]
\frac{1}{N_c} \tr\left[\Omega^{(-k)}(0)\right]>.
\eeeq

By repeating the procedure we have followed in computing the partition function,
keeping again the projection onto the $m_c$ sector in the decomposition $U(N_c)=U(1)
\times SU(N_c)/Z_{N_c}$ and subtracting the classical background, we get \cite{gsv}

\beeq
\label{correp}
&&\frac{1}{N_c^2} <\tr\left[\Omega^{(k)}(x)\right] \tr\left[\Omega^{(-k)}(0)\right]>
=\frac{1}{{\cal Z}N_c!}\sqrt{\frac{2\pi}{{\cal A}N_c}}\exp\left[
-\frac{k^2 x {\cal A}}{4\pi R_c}\left(1-\frac{x}{2\pi R_c N_c}\right)\right]
\no \\ \nonumber
&&\times\sum_{n_i\neq n_j}
\exp\left[-\frac{{\cal A}}{2}\left(\sum_{i=1}^{N_c} n_i^2-\frac{1}{N_c}
\left(\sum_{i=1}^{N_c}n_i\right)^2\right)\right]\frac{1}{N_c}
\sum_{j=1}^{N_c}\exp\left[-\frac{xk{\cal A}}{2\pi R_c}\left(n_j-
\frac{1}{N_c}\sum_{i=1}^{N_c}n_i\right)\right] \\ 
&&\times\int_0^{2\pi}\frac{d\alpha}{\sqrt {\pi}} \exp\left[-\left(\alpha -\frac{2\pi}{N_c}
\sum_{i=1}^{N_c}n_i\right)^2 -2\pi im_c\left(
\frac{N_c-1}{2}- \frac{1}{N_c}
\sum_{i=1}^{N_c}n_i\right)\right].
\eeeq

Eqs.~\re{final}, \re{correp} entail $N_c$ sums over the different integers
$n_i$ which can take any value between $-\infty$ and $+\infty$. As a
consequence of the $SU(N_c)/Z_{N_c}$ symmetry, those equations are manifestly
invariant under a simultaneous shift of all the $n_i$ by an integer. 

We start by considering large $N$ values in order to comply
with our perturbative treatment; this forces even larger
values for $N_c$ \cite{bv}.

Obviously, plenty of different configurations are possible and to sum
over all of them is beyond  reach. We are therefore seeking for 
configurations which may be dominant in particular physical regimes.
In a recent paper \cite{mrw}, three basic different regimes have been presented
for a scalar noncommutative theory in two dimensions, when approximated
by means of a $M\times M$ matrix model. Three different phases (disordered,
planar and GMS \cite{gms} ) are possible, according to the behaviour of the
noncommutativity parameter $\theta$ with respect to the integer $M$
which is to be sent  
eventually to $\infty$ ($\theta \sim M^{\nu}$ with $\nu <1$, $\n=1$, $\n>1$, 
respectively).
This integer in turn is related to a large distance cutoff $L$ of the theory,
which can be identified in our case with the length of the side of the square torus.
Moreover in a $U(N)$ gauge theory, $M$ cannot be smaller than $N$.

Eqs.~\re{final}, \re{correp} exhibit a Gaussian damping with
respect to the ``occupation numbers'' $n_i$, which
suggests that a kind of saddle-point approximation may be feasible. 
The most favoured configurations
are those with minimal fluctuations. This
happens when the integers $n_i$  assume adjacent values.

In the saddle-point approximation, for large values of $N_c$,  we eventually get 
\cite{bv}
\beq
\label{corref1}
{\cal W}_2(k)\equiv <W(k)W(-k)>\simeq
\frac{\exp\left[\frac{{\cal A}}{2}|k|(N_c-1-|k|)\right]-1}
{\frac{{\cal A}}{2}N_c|k|(N_c-1-|k|)}. 
\eeq

Remarkably, when $|k|<N_c-1$, we find a correlation function exponentially
{\it increasing} with $|k|$, in qualitative agreement with an analogous finding 
in \cite{gross} and with our perturbative result. 
In order to reach a quantitative agreement, 
a fine tuning of the exponents is possible and entails a relation of the
kind $\theta \sim R^{\nu}$ with $\nu \le 1$ (see \cite{bv}).

If we remember that $R$ is the
natural cutoff of our formulation, the condition $\theta \sim R^{\nu}$ with
$\nu \leq 1$ is reminiscent of the analogous condition in \cite{mrw}  with respect
to the cutoff $M$ (or the torus side length $L$), 
related to  the dimension of their matrix model. 
Actually $\nu <1$ describes the disordered phase, where quantum effects are
dominant, while $\nu =1$ is a border-line value, related to the so-called
planar phase. The values $\nu >1$ would correspond to the GMS phase which
is inaccessible to our treatment. 

As a final remark we notice that Eq.~\re{corref1} changes dramatically 
when $|k| > N_c-1$, strongly deviating from the perturbative result
and thus possibly suggesting the onset of a new phase. Nevertheless,
we ought to recall that in this region the saddle-point approximation
we adopted no longer holds \cite{bv}.

\vskip1cm
\noindent

\section*{ACKNOWLEDGEMENTS}

We thank G. De Pol and A. Torrielli for useful discussions.

\end{document}